\documentclass[preprint]{elsarticle}
\usepackage{amsmath}
\usepackage{amssymb}
\usepackage{bm}
\usepackage{mathrsfs}
\begin{document}

\title{Studies on multiplication effect of noises of PPDs, and a proposal of a new structure to improve the performance}
\author[phys]{H.~Oide\corref{cor1}}
\ead{oide@icepp.s.u-tokyo.ac.jp}
\author[phys]{T.~Murase}
\author[phys]{H.~Otono}
\author[icepp]{S.~Yamashita}

\cortext[cor1]{Corresponding author}

\address[phys]{Department of Physics, University of Tokyo, 7-3-1, Hongo, Bunkyo, Tokyo, Japan}
\address[icepp]{ICEPP, University of Tokyo, 7-3-1, Hongo, Bunkyo, Tokyo, Japan}

\date{}

\begin{abstract}
Pixelated Photon Detectors (PPDs) are the most promising semiconductor photodetectors in recent years.
One of the issues with the PPD is its high noise rate.
As well as random noise, PPD also exhibits so called {\it after-pulsing} and {\it optical crosstalk}, and these limit the applicable range of its gain as well as its size.
By accurately measuring each of these causes of noises independently, we quantitatively evaluated how the performance of the present device is limited by multiplication effect of these noises. 
With this result and the pulsing mechanism of PPD, we propose a new structure of PPD which could have high gain with low noise.
\end{abstract}


\maketitle

\section{Introduction}
Pixelated Photon Detectors (PPDs), also known as G-APD, MPPC, SiPM, SPAD, etc., are semiconductor photodetectors in which APDs operated in Geiger-mode are pixelated in an array\cite{Renker}.
Each pixel of a PPD is a binary detector with an avalanche multiplication factor (gain) typically of the order of $10^{5}-10^{6}$.
They are also capable of single photon counting.
As PPDs have many advantages over conventional photomultiplier tubes (PMTs) such as high photon detection efficiency, magnetic field tolerance, low operational voltage, ultra-thin body, etc., they are one of the most promising of the next generation of photodetectors. 
Nevertheless the gain of present PPDs is somewhat less than that of PMTs, thus increasing their gain is desirable. Other areas for improvement for PPDs among others are noise reduction, increase of effective area, and improvement of radiation hardness.

As PPDs are operated in Geiger-mode, some quenching mechanism is mandatory to terminate the avalanche multiplication.
One such quenching mechanism is {\it passive quenching}, in which a resistor with high resistance is connected in series to the diode of each pixel. 
It is known that for passively quenched PPDs, the gain $G$ can be modeled by
\begin{eqnarray}
\label{eq1}
G=C_{\rm d}(V_{\rm op}-V_{0})/e
\end{eqnarray}
where $V_{\rm op}$, $V_{0}$, $C_{\rm d}$, and $e$ denote the operational voltage, the breakdown voltage, the capacitance of each pixel, and elementary electric charge, respectively. The breakdown voltage is the threshold in which Geiger-mode gets started.
$V_{0}$ is determined by the concentration distribution of impurities inside the diode and the impact ionization probability of electrons and holes in Si. The above relation is generally valid for passively quenched PPDs independent of the particular instantiation\cite{Otono}.
As can be seen from the equation above, the simplest way to accomplish higher gain is to increase the over-voltage $\Delta V\equiv V_{\rm op}-V_{0}$, but such an approach is of limited practicality due to a corresponding increase in the frequency of noise pulses.
\begin{figure}[btp]
\begin{center}
\includegraphics[width=80mm]{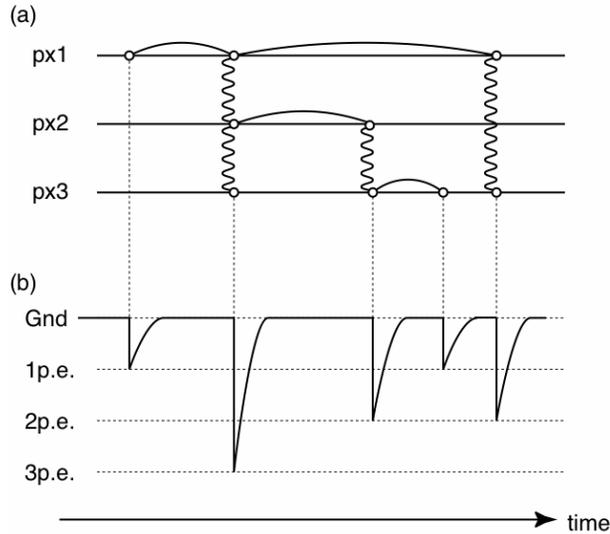}
\caption{(a): A schematic showing how continuous noise pulsing can arise from a single pulse (b): The observed waveform of the pulsing of (a).}
\label{diagram}
\end{center}
\end{figure}

\begin{figure}[tbp]
\begin{center}
\includegraphics[width=80mm]{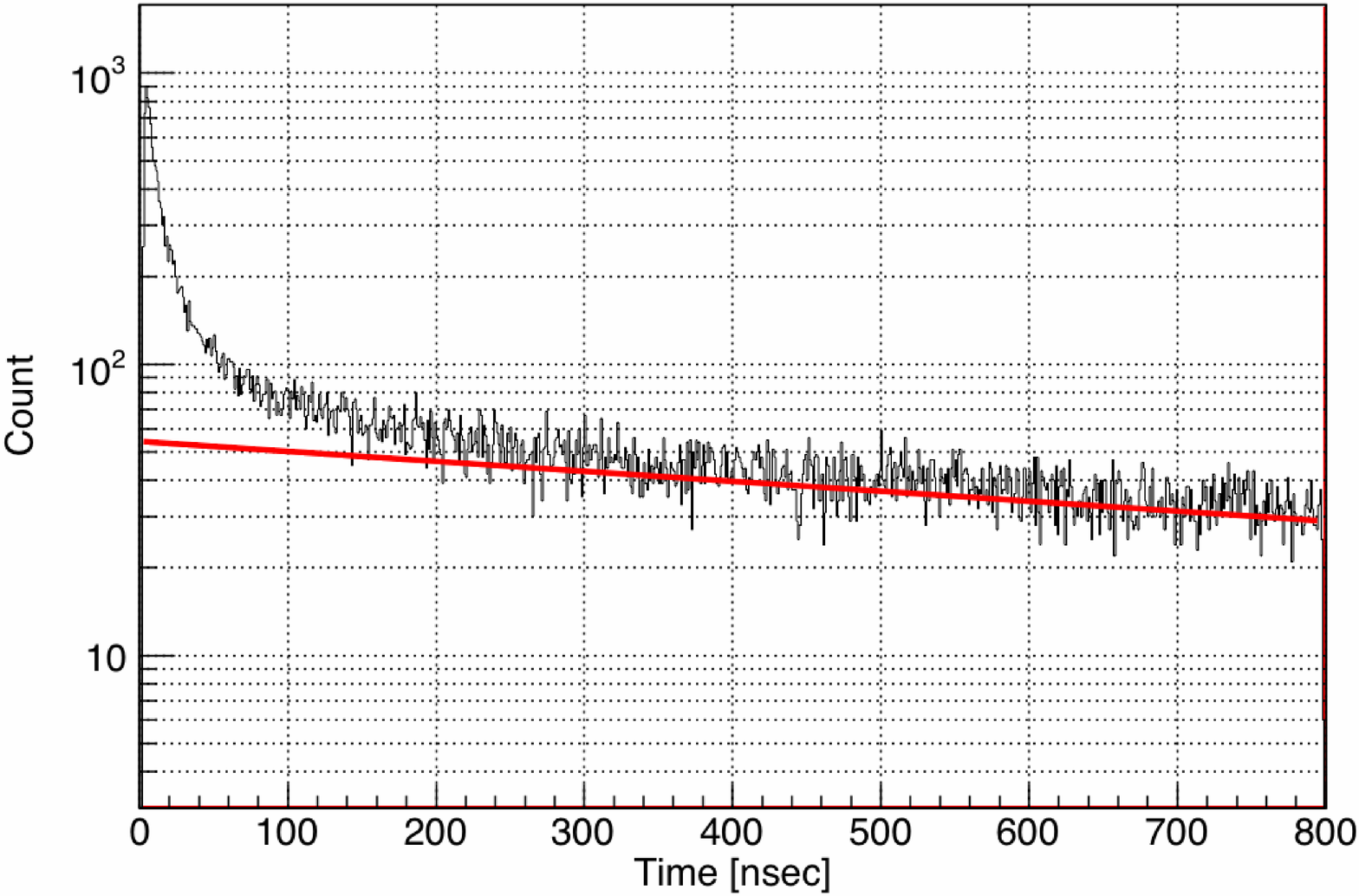}
\caption{The time interval distribution at $\Delta V=3.45{\rm V}$. The line represents the distribution of random noise.}
\label{ap dist}
\end{center}

\begin{center}
\includegraphics[width=80mm]{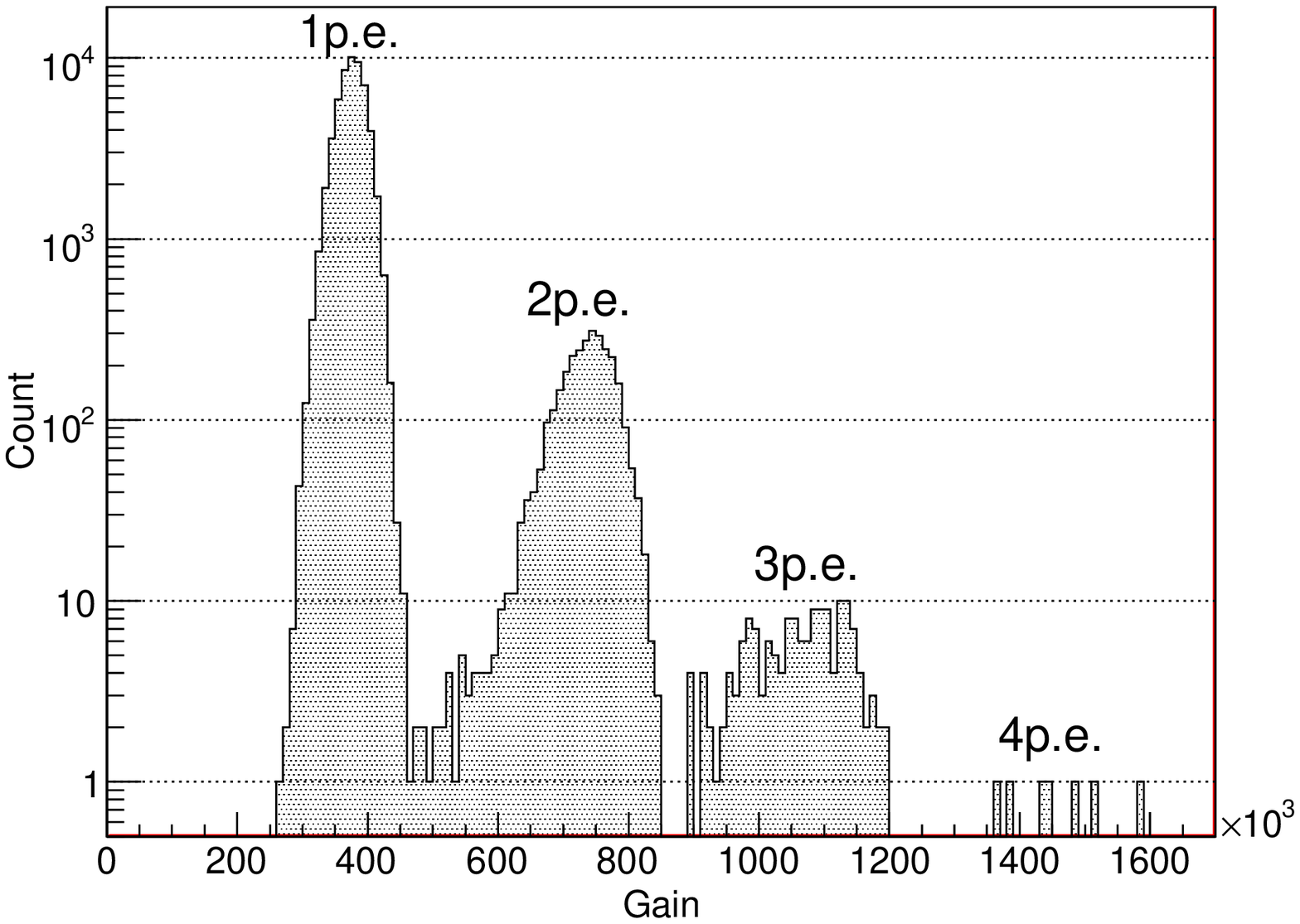}
\caption{The distribution of the magnitude of pulses at $\Delta V=2.65{\rm V}$. Note that the X-axis is converted to gain.}
\label{gain}
\end{center}
\end{figure}

In a PPD, Not only photoelectrons but also free carriers (electrons and holes) inside the lattice generate noise pulses of gain $10^{5}-10^{6}$.
Noises of PPDs can be divided into three categories:  random pulsing, after-pulsing, and optical crosstalk \cite{Buzhan,Sciacca,Rech}.
After-pulsing and optical crosstalk (crosstalk) are both subordinary pulsing incident with pulses from photoelectron signal or random pulse noise.
The noise rate of PPDs is typically a few hundred kcps per ${\rm mm^{2}}$ when the operational voltage is in the ``appropriate region''. An increase of the voltage above this region leads to a drastic increase in the noise rate to a few Mcps or higher.
This is due to a successive and continuous generation of noise pulses because of an increase in the probability of after-pulsing and crosstalk (see Fig.\ref{diagram}).
To investigate this phenomenon, we measured separately the dependencies of random pulsing, after-pulsing, and crosstalk on $\Delta V$, then reconstructed the increase in the total-noise rate from the measured relations.

In this paper we report the results of these studies and speculated mechanism inside PPDs from them.
We also discuss the possibilities for increasing gain while suppressing noises based on our study of the internal mechanics of PPDs.

\section{Measurements: the methods}
The PPD we measured in this paper is a Multi-pixel Photon Detector (MPPC) made by Hamamatsu Photonics K.K., which consists of 1600 pixels of Geiger-mode APD, each of area $25\times25{\rm \mu m^{2}}$ (type {\tt S10362-11-25})\cite{HPK, MPPC}.
A Hamamatsu {\tt C5594} voltage amplifier was used for readout. Its gain is 36dB (63-fold voltage amplification) and its bandwidth is 1.5GHz.
The quantities we wished to measure were the random-noise rate, total-noise rate, after-pulsing probability, crosstalk probability, and relative photon detection efficiency (PDE).
All measurements were performed at room temperature (300K).
The breakdown voltage of the sample we used is 73.5V at this temperature.
All measurements were performed in a dark chamber, and no light source was used except in the measurement of the relative PDE.
The operational voltage range we used was from $\Delta V=1.5{\rm V}$ to $\Delta V=5.1{\rm V}$.
Other operational voltage regions were excluded from from our study for the following reasons: in the range $\Delta V<1.5{\rm V}$ the signal from the readout amplifier was smaller than the amplifier's noise, and in the range $\Delta V>5.1{\rm V}$ the high frequency of noise pulses precluded the use of our waveform analysis (referred later).
The error in the operational voltage for each measured point was $\pm0.04{\rm V}$, which includes errors arising from the stability of the HV supply and temperature fluctuations of the environment.

\subsection{After-pulsing probability, random-noise rate}
The method for measuring after-pulsing is to measure the time interval between a particular pulse and its next pulse in the dark chamber, and to make a distribution of this interval.
If there is no after-pulsing, the distribution decreases exponentially and the inverse of the time constant of the distribution $\tau_{\rm N}$ is the rate of random pulsing (random-noise rate).
When after-pulsing exists, an {\it afterpulse} tends to be emitted close in time to the pulse which generates it so that the time constant(s) of after-pulsing is relatively short\cite{Cova}. Afterpulses with shorter time constant(s) than that of random pulsing are observed as an excess in the distribution of random pulsing.
Acquiring waveforms with a digital oscilloscope and applying offline waveform analyses, we measured this distribution\cite{NDIP}.
An example of a distribution obtained by this procedure is shown in Fig.\ref{ap dist}. 
Here, the long tail component is from random pulsing, while the excess seen above this distribution at short times is from after-pulsing.
After-pulsing has several classes each corresponding to a separate time constant. The time interval distribution can be fitted with a function
\begin{eqnarray}
n(t;\Delta V)=\sum_{j}A_{j}\exp(-t/\tau_{j})+A_{\rm N}\exp(-t/\tau_{\rm N})\nonumber\\~~(j=1,2,\cdots)~.
\end{eqnarray}
where $\tau_{j}$ denotes each time constant of after-pulsing, and $1/\tau_{\rm N}$ denotes the random-noise rate. $A_{j}$ and $A_{\rm N}$ denote the coefficient corresponding to the time constant of the same index.
Assuming that the after-pulsing distribution has two components, the fitted time constants were $8.6\pm2{\rm ns}$ and $74^{+50}_{-20}{\rm ns}$.
No significant operational voltage dependence was found for each time constant, neither was there one for the ratio of the two coefficients of $A_{1}/A_{2}$.
Thus we set the two time constants $\tau_{1},~\tau_{2}$ to $8.6{\rm ns}$ and $74{\rm ns}$ respectively, and fixed the ratio $A_{2}/A_{1}$ to a constant $c$ for all measurement voltage points.
Then a fitting was performed for the two remaining parameters $A_{1}$ and $A_{N}$.
As a result a quantity $P_{\rm AP}(\Delta V)$ is obtained for a given over-voltage:
\begin{eqnarray}
\label{eq:approb}
P_{\rm AP}(\Delta V)\equiv\frac{A_{1}\tau_{1}+cA_{1}\tau_{2}}{A_{1}\tau_{1}+cA_{1}\tau_{2}+A_{\rm N}\tau_{\rm N}}
\end{eqnarray}
We defined this quantity as the after-pulsing probability at that voltage.

With this distribution (Fig.\ref{ap dist}) it can be seen that by counting the number of noise pulses after a veto time of about $1{\rm \mu s}$, we can measure the random-noise rate separately.
Note that the random-noise rate is corrected for accidental coincidence of random pulses and the veto time.

\subsection{Total-noise rate}
\label{sec:totnoise}
On the other hand, following a similar procedure as for the random noise rate measurement, but minimizing the veto time gives us the total-noise rate including afterpulses.
Our measurement system had a dead time of 13.5ns after triggering from a pulse.
As shown in Fig.\ref{ap dist}, an afterpulse is correlated to its foregoing pulse.
This nature of after-pulsing needs a special treatment to calibrate the dead time after triggering.
The detail of this treatment is described in \ref{sec:reconstruction}.

\subsection{Crosstalk probability}
Crosstalk occurs when the photons emitted during an avalanche multiplication of a certain pixel propagate to other pixels.
This results in an avalanche in these other pixels thus resulting in a number of pixels firing at the same time.
The probability of crosstalk generation for a random-noise pulse is measured by investigating the distribution of the magnitude of pulses in a dark chamber.
When the (total) noise rate is high, pile-up of pulses is a problem when measuring the magnitude of each pulse.
Using the waveform analysis method mentioned above, we succeeded in accurately measuring the magnitude of an individual pulse, excluding the influence of its neighboring pulses.
Fig.\ref{gain} shows an example of the magnitude distribution measured in this way.
Several peaks are found, and they are labelled 1p.e., 2p.e., $\cdots$, in order of magnitude.
We define the ratio of pulses over 2p.e. to all pulses in the distribution (over-2p.e.) as crosstalk probability.
We also measured the ratio of pulses over 3p.e. to all pulses (over-3p.e.) as a reference.

\subsection{Relative PDE}
A simple measurement of the relative PDE (photon detection efficiency) was made to be used for later discussion.
By comparing the cases of turning an LED on and off within the dark chamber, the voltage dependence of PDE was measured.
As a light source a blue LED of $470{\rm nm}$ wavelength was used, and light emitted from the LED was diffused sufficiently so that mostly only single photon reached the PPD.
Measured PDE values were normalized to the value at $V_{\rm op}=75.0{\rm V}~(\Delta V=1.55{\rm V})$.

\begin{figure}[tbp]
\begin{center}
\includegraphics[width=80mm]{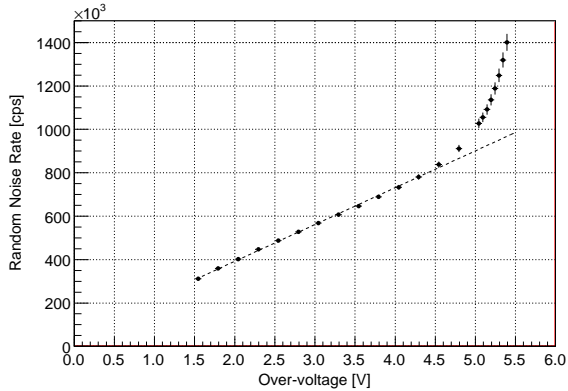}
\caption{The $\Delta V$-dependence of random-noise rate. It is considered that the rapid increase at $\Delta V\gtrsim4.5{\rm V}$ is the influence of leakage of afterpulses from the veto gate.}
\label{noise rate}
\end{center}
\end{figure}

\begin{figure}
\begin{center}
\includegraphics[width=80mm]{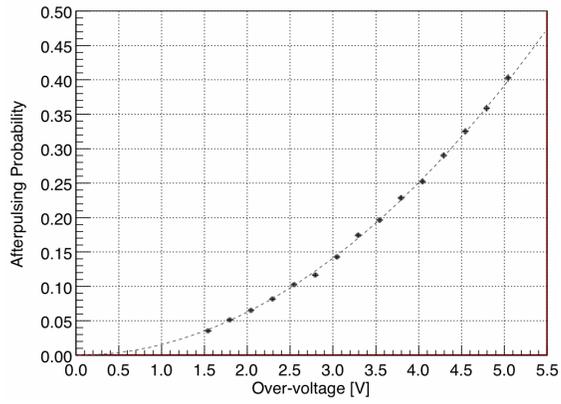}
\caption{The $\Delta V$-dependence of after-pulsing probability. The dotted line represents the fit of the measurement points with a function that is proportional to $\Delta V^{2}$.}
\label{afterpulse}
\end{center}
\end{figure}

\begin{figure}
\begin{center}
\includegraphics[width=80mm]{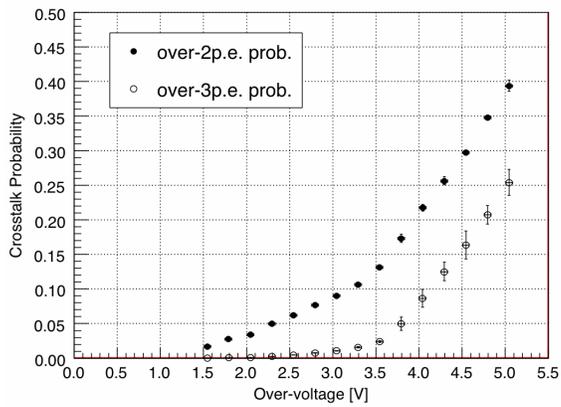}
\caption{The $\Delta V$-dependence of crosstalk probability}
\label{crosstalk}
\end{center}
\end{figure}

\section{Measurements: the results}
\subsection{About each measurement}
\subsubsection*{\underline{Random-noise rate}}
Fig.\ref{noise rate} shows the voltage dependence of the random-noise rate.
In the range $1.5{\rm V}\lesssim\Delta V\lesssim4.5{\rm V}$, random pulsing increases linearly with $\Delta V$.
On the other hand a large excess from linearity is observed at $\Delta V\gtrsim4.5{\rm V}$.
This excess is considered not to be the real increase of random pulsing itself, but to be the effect of afterpulse leakage over the $1{\rm \mu s}$ veto.
Further discussions on this issue are presented in the next section.

\subsubsection*{\underline{After-pulsing probability}}
The result of the measurement of the voltage dependence of after-pulsing probability is shown in Fig.\ref{afterpulse}.
The dotted line is the $\chi^{2}$ fit of measured points by the function that is proportional to $\Delta V^{2}$.
There is fairly good agreement between the measurements and the $\chi^{2}$ fit.
This proportionality to $\Delta V^{2}$ is also valid for each of after-pulsing belonging to the two time constants $\tau_{1}$ and $\tau_{2}$, since the ratio $A_{1}/A_{2}$ is constant with over-voltages.
Thus after-pulsing probability including its time dependence is expressed as
\begin{eqnarray}
{\mathscr P}_{\rm AP}(t;\Delta V)=\alpha_{1}\Delta V^{2}e^{-t/\tau_{1}}+\alpha_{2}\Delta V^{2}e^{-t/\tau_{2}}~~.
\end{eqnarray}

\subsubsection*{\underline{Crosstalk probability}}
Similarly Fig.\ref{crosstalk} shows the voltage dependence of over-2p.e. crosstalk probability and of over-3p.e. crosstalk probability.
In contrast to after-pulsing, the dependence of crosstalk probability on $\Delta V$ is more complex.
In particular there is a kink in over-2p.e. crosstalk probability at $\Delta V\simeq3.5{\rm V}$.
Also, we can see a rapid increase in over-3p.e. at $\Delta V\gtrsim3.5{\rm V}$.
Thus the 2p.e. kink indicates the increase of over-3p.e. crosstalk probability.

\subsubsection*{\underline{Relative PDE}}
\begin{figure}[t]
\begin{center}
\includegraphics[width=80mm]{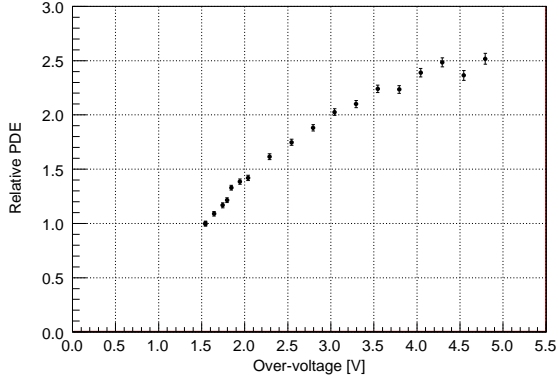}
\caption{The voltage dependence of relative PDEDNote that each value is normalized to that of $V_{\rm op}=75.0{\rm V}$~$(\Delta V=1.55{\rm V})$.}
\label{pde}
\end{center}
\end{figure}

Fig.\ref{pde} shows the result of the voltage dependence of relative PDE.
With this result and the fact that PDE should be zero at $\Delta V=0.0{\rm V}$, it is observed that PDE tends to saturate as $\Delta V$ increases.
Similar results can be seen for example in \cite{Uozumi}.

\subsection{Reconstruction of total-noise rate}
\label{sec:reconstruction}
As described above, random-noise rate, after-pulsing and crosstalk probabilities were measured for each operational voltage point.
Here we first derive a model to reconstruct the total-noise rate from these measurements.
Next we describe a method to calibrate the results of a total-noise rate measurement, as mentioned in \ref{sec:totnoise}.
Finally the reconstructed model and the measured total-noise rate are compared.

At first, we describe the model which reconstructs the total-noise rate from random-noise rate using the after-pulsing and crosstalk probabilities we measured.
To obtain a full understanding of the multiplication effect of after-pulsing and crosstalk on each other, some full simulation, e.g. a MC method, is needed.
However, to know only the mean of the total-noise rate in time, this is not necessary.

When a particular pulse is generated, generally several pixels fire signals at the same time due to crosstalk.
Pulses from each of the fired pixels sum to form the signal.
The magnitude of the signal contains information on the number of pixels which fired.
The magnitude of pulses has a distribution like Fig.\ref{gain}.
For the case of neglecting after-pulsing, the observed number of signals originating from a particular pulse is identically unity:
\begin{eqnarray}
\label{eq5}
X_{0}=1\times\bm{1}=\bm{1}~.
\end{eqnarray}
$X_{k}$ denotes the mean number of observed signals which derive from a particular pulse. The index $k$ means that up to $k$ afterpulses are considered for each pixel.
(In Eq.(\ref{eq5}) no afterpulse is considered so that the index of $X$ is zero.)
The plain letter ``1'' denotes the probability of observing just {\it one signal} arising from a particular pulse, and the bold letter ``$\bm{1}$'' denotes this ``one signal''.

We define $q_{j}$ as the probability that $j$ pixels fire at the same time for a particular pulse because of crosstalk.
Note that $\sum_{j=1}q_{j}=1$.
With this notation, over-2p.e. crosstalk probability is $1-q_{1}$, and over-3p.e. crosstalk probability is $1-(q_{1}+q_{2})$.

Now consider the case where up to 1 afterpulse arises for each pixel (up to 1st order after-pulsing).
We initially let $j$ pixels fire.
We then consider after-pulses for each of the $j$ pixels individually.

Now let $p$ be the after-pulsing probability $P_{\rm AP}(\Delta V)$ at a given $\Delta V$.
Then the mean number of afterpulses observed is:
\begin{eqnarray}
\label{eq:5}
q_{1}\times\bm{1}\times p+q_{2}\times\bm{2}\times p+\cdots=\sum_{j}\bm{j}q_{j}p=\bar{\bm{q}}p~~.
\end{eqnarray}
where $\bar{\bm{q}}\equiv\sum_{j=1}\bm{j}q_{j}~$.
Adding $\bm{1}$ for the first pulses, $X_{1}$ is obtained:
\begin{eqnarray}
X_{1}&=&\bm{1}+\bar{\bm{q}}p~.
\end{eqnarray}
Considering up to 2nd order after-pulsing, $X_{2}$ is
\begin{eqnarray}
X_{2}&=&\bm{1}+\bar{\bm{q}}p(\bm{1}+\bar{\bm{q}}p)=\bm{1}+\bar{\bm{q}}p+(\bar{\bm{q}}p)^{2}~~.
\end{eqnarray}
Similarly considering up to infinity-order after-pulsing,
\begin{eqnarray}
X_{\infty}=\bm{1}+(\bar{\bm{q}}p)+(\bar{\bm{q}}p)^{2}+\cdots=\frac{1}{1-\bar{q}p}\equiv\xi~.
\end{eqnarray}
Thus the total-noise rate $N_{\rm tot}$ is expressed as
\begin{eqnarray}
\label{totnoise}
N_{\rm tot}=\xi N_{\rm rand}=\frac{1}{1-\bar{q}p}N_{\rm rand},
\end{eqnarray}
where $N_{\rm rand}$ denotes the random-noise rate.
Note that we assumed that the random-noise rate $N_{\rm rand}$ increases linearly in the region $\Delta V\gtrsim4.5{\rm V}$, as is the case for the lower region.
Eq.(\ref{totnoise}) represents a model for the total-noise rate which is reconstructed from the random-noise rate including the influence of after-pulsing and crosstalk.
Hereafter we call $N_{\rm tot}$ ``reconstructed noise rate''.
Note that $N_{\rm tot}$ diverges at $\bar{q}p=1$.

Next we look at how to calibrate the measured total-noise rate.
As we have described in \ref{sec:totnoise}, the measurement system we used had a dead time of $T_{d}=13.5{\rm ns}$.
In addition to the calibration of random pulsing in the dead time, calibration of after-pulsing is necessary, since afterpulses are correlated to the initial pulse which generates them.
Considering up to 1st order after-pulsing, the after-pulsing calibration is expressed only by a quantity $\beta$:
\begin{eqnarray}
\beta&\equiv&\frac{\displaystyle{\int_{T_{\rm d}}^{\infty}\!dt\,(\alpha_{1}\Delta V^{2}e^{-t/\tau_{1}}+\alpha_{2}\Delta V^{2}e^{-t/\tau_{2}})}}{\displaystyle{\int_{0}^{\infty}\!dt\,(\alpha_{1}\Delta V^{2}e^{-t/\tau_{1}}+\alpha_{2}\Delta V^{2}e^{-t/\tau_{2}})}}\nonumber\\
&=&\frac{\alpha_{1}\tau_{1}e^{-T_{\rm d}/\tau_{1}}+\alpha_{2}\tau_{2}e^{-T_{\rm d}/\tau_{2}}}{\alpha_{1}\tau_{1}+\alpha_{2}\tau_{2}}~,
\end{eqnarray}
where $\beta$ is a voltage-independent constant\footnote{The independence of $\beta$ from $\Delta V$ comes from the proportionality of ${\mathscr P}_{\rm AP}(t;\Delta V)$ to $\Delta V^{2}$.}.
Define $N_{\rm m}$ as the total-noise rate which is calibrated only for accidental coincidence of the dead time with random pulses.
Then the total-noise rate which is calibrated also for after-pulsing $N_{\rm m}^{\rm (c)}$ is
\begin{eqnarray}
N_{\rm m}^{\rm (c)}=\frac{1}{1-(1-\beta)\bar{q}p}N_{\rm rand}~.
\end{eqnarray}
We set this $N_{\rm m}^{\rm (c)}$ as the measured total-noise rate to be compared with the reconstructed noise rate.
Note that for a calibration which considers 2nd or higher order after-pulsing, the calculation becomes quite complex as the time origin of after-pulsing varies.
In this paper we carried out the calibration only considering the 1st order after-pulsing.

\begin{figure}[tbp]
\begin{center}
\includegraphics[width=90mm]{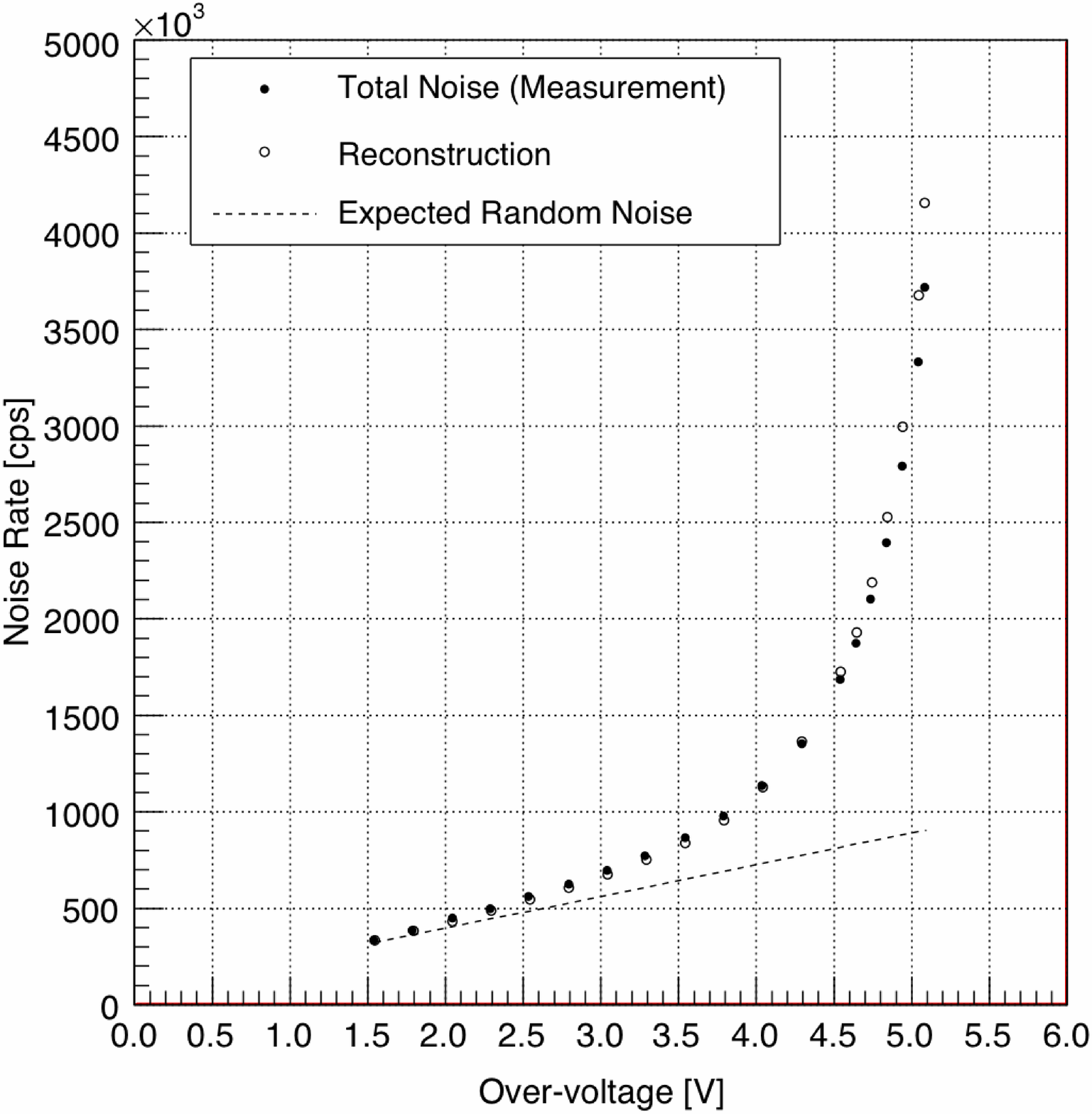}
\caption{Reconstruction and the measurement of total-noise rate.}
\label{reconst}
\end{center}
\end{figure}

Fig.\ref{reconst} shows the comparison of the reconstructed noise rate and the measured one.
Good agreement between the reconstruction and measurement is seen.
The residue gets slightly larger as $\Delta V$ increases past $\Delta V\gtrsim4.5{\rm V}$.
This is possibly due to the effect of 2nd or higher order after-pulsing.
The increase of total-noise rate with $\Delta V$ is successfully deduced from the voltage dependence of random pulsing, after-pulsing and crosstalk probability.

\section{Discussion}
\subsection{Random-noise rate and PDE}
First we consider the voltage dependence of the random-noise rate.
The random-noise rate is determined by the product of the number of carriers generated randomly and the {\it avalanche probability}.
Here we define ``avalanche probability'' as the probability that a particular carrier generates a Geiger-mode avalanche multiplication.
At 300K the number of carriers generated in the lattice is mainly determined by thermal excitation\cite{Grundmann}.
Since the variation of the electric field in the measured region is $\sim5\%$, we consider that the effect of the electric field on the carrier number is negligible.
Thus the proportionality of random-noise rate with $\Delta V$ cannot be explained by carriers induced by the electric field, since the PPDs enters Geiger-mode when $V_{\rm op}>V_{0}$.
Therefore we presume that the $\Delta V$-dependence of the random-noise rate reflects the $\Delta V$-dependence of the avalanche probability.

Next we consider the voltage dependence of the PDE.
The PDE is determined by the product of the quantum efficiency (QE) of photoelectric conversion, the acceptance of the device and the avalanche probability.
QE and the acceptance are not dependent on operational voltage.
Thus we assume that the voltage dependence of the PDE should also reflect the voltage dependence of the avalanche probability.

If the random-noise rate and the PDE both reflect the {\it same} avalanche probability, both of them must show a similar voltage dependence.
However, as shown in Fig.\ref{noise rate} and Fig.\ref{pde}, the random-noise rate increases linearly with $\Delta V$, while the PDE tends to saturate as $\Delta V$ increases.

Avalanche multiplication arises from impact ionization of electrons and holes, but the impact ionization rate of electrons is about an order of magnitude larger than that of holes\cite{II}.
It is considered that the probability of avalanche multiplication which originates from an electron should be different to that originating from a hole.

\begin{figure}[tbp]
\begin{center}
\includegraphics[width=80mm]{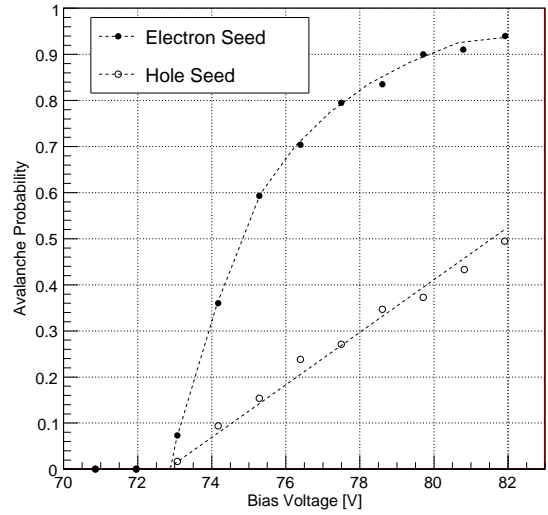}
\caption{The voltage dependence of avalanche probability of electrons and holes based on the simulation. Note that the breakdown voltage varies with the width of the depletion layer, but the $\Delta V$-dependence of the avalanche probability does not change drastically.}
\label{iimc}
\end{center}
\end{figure}

\begin{figure}[tbp]
\begin{center}
\includegraphics[width=80mm]{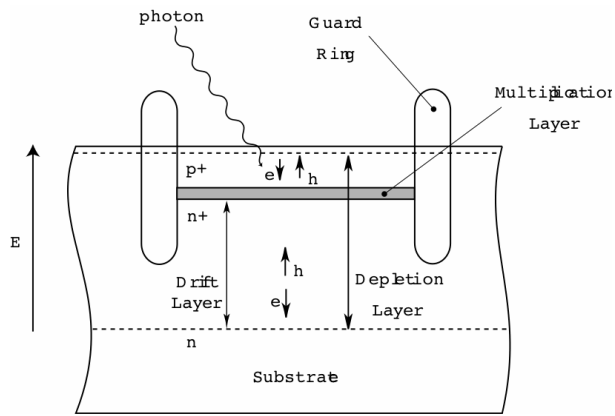}
\caption{The structure of a p-on-n type PPD.}
\label{structure}
\end{center}
\end{figure}

To investigate the relation between the difference of the $\Delta V$-dependence of the random-noise rate and the PDE, and the difference of the avalanche probability for an electron and a hole, we performed a simple simulation as follows:
Consider a one dimensional region in the $z$ direction in which there is a gaussian-like electric field which has a width of the order of ${\rm \mu m}$. This width corresponds to the depth of the depletion layer.
Assuming that the depletion layer does not change in $V_{\rm op}$, the magnitude of the electric field at each point in $z$ is proportional to $V_{\rm op}$ which is identical to the integral of the entire field.
The impact ionization rate per unit length at a given electric field of electrons and holes $\alpha_{\rm e},~\alpha_{\rm h} {\rm [1/cm]}$, respectively, can be written by:
\begin{eqnarray}
\alpha_{\rm e}(E)=\exp\left(A_{\rm e}+\frac{B_{\rm e}}{E}\right)\\
\alpha_{\rm h}(E)=\exp\left(A_{\rm h}+\frac{B_{\rm h}}{E}\right)
\end{eqnarray}
where $E$ denotes the electric field in the direction of $z$ which is expressed in ${\rm [V/cm]}$, and $A_{\rm e}=1.35\times10,~A_{\rm h}=1.44\times10,$ and $B_{\rm e}=-1.17\times10^{6}~{\rm [V/cm]},~B_{\rm h}=-1.95\times10^{6}~{\rm [V/cm]}$\cite{II}.
At a high electric field such that PPDs are in Geiger-mode, the drift speeds of electrons and holes are sufficiently saturated to a speed of $10^{7}{\rm cm/s}$.
Thus for the case of the speed we simply placed the electrons with the saturation speed at a point of low electric potential where the impact ionization probability is negligible, and we propagated the electrons into the high field region.
Similarly for the holes for which the initial position was set to a point of high potential.

Each multiplication process is simulated then the simulation is stopped after a given time.
If the number of electron-hole pairs is greater than $10^{4}$ then the event is categorized as Geiger-mode accomplished, otherwise not.
Note that the loss of carriers from recombination, etc. is not considered.
The result of this simulation is shown in Fig.\ref{iimc}.

Summarizing the result, if the original carrier is an electron then the avalanche probability tends to saturate as $\Delta V$ increases; if the original carrier is a hole then the avalanche probability is proportional to $\Delta V$.
By assuming that the random-noise rate reflects the avalanche probability originating from holes and that the PDE reflects the avalanche probability originating from electrons, it is possible to account for the $\Delta V$-dependence for each.

Hereafter we discuss qualitatively the validity of this assertion.
The PPD that we used in this paper is a p-on-n type (see Fig.\ref{structure}).
In addition, since the PPD has a sensitivity to photons of 470nm wavelength for which the absorption length in Si is about $0.5{\rm \mu m}$, the multiplication layer should be relatively close to the incidence surface\cite{Dash,Piemonte}.
The photoelectric conversion would mainly occur above the depletion layer and the generated electron would then propagate to the multiplication layer.
On the other hand, most of the depletion layer is below the multiplication layer so that for random (thermal) excitations of electron-hole pairs, mainly holes drift to the multiplication layer.
This excitation would not only happen in the drift layer of the depletion layer but also in the substrate.
Hence it is possible to account for the difference of the dependencies of the random-noise rate and the PDE on $\Delta V$.

It can be seen thus that a p-on-n type PPD inherently suppresses noises: it enlarges the avalanche probability for incident photons and suppresses the avalanche probability for random excitations.
MPPCs have the structure described above.

\subsection{After-pulsing}
Next we discuss the $\Delta V^{2}$ dependence of the after-pulsing probability.
The origin of after-pulsing is supposed to be the trapping and re-release of carriers at lattice defects.
It is natural to expect that the after-pulsing probability is determined by the product of the initial number of carriers, the trap (and re-release) probability, and avalanche probability.

The number of carriers is proportional to $\Delta V$ (see Eq.(\ref{eq1})), while the trap (and re-release) probability is not strongly dependent on the electric field.
Thus the avalanche probability should be proportional to $\Delta V$ to account for the voltage dependence of the after-pulsing probability.
This proportionality of avalanche probability to $\Delta V$ suggests that the avalanche multiplication originating from holes is the main source of after-pulsing.
A description of a process that accounts for this dependency is as follows: free electrons generated at multiplication layer are captured in the drift layer below the multiplication layer, and holes released from traps drift to the multiplication layer and resulting in an afterpulse.
Further investigation is needed to confirm that the phenomenon described is in fact the cause of after-pulsing.

\subsection{The effect of noise reduction}
The reconstruction of the total-noise rate described above reduces the total-noise rate to the random-noise rate, the after-pulsing and crosstalk probabilities.
Thus it is possible to estimate the total-noise rate from a given voltage dependence of these factors.
This enables a numerical calculation of the noise rate which is useful in the development of PPDs.

\begin{figure}[tbp]
\begin{center}
\includegraphics[width=90mm]{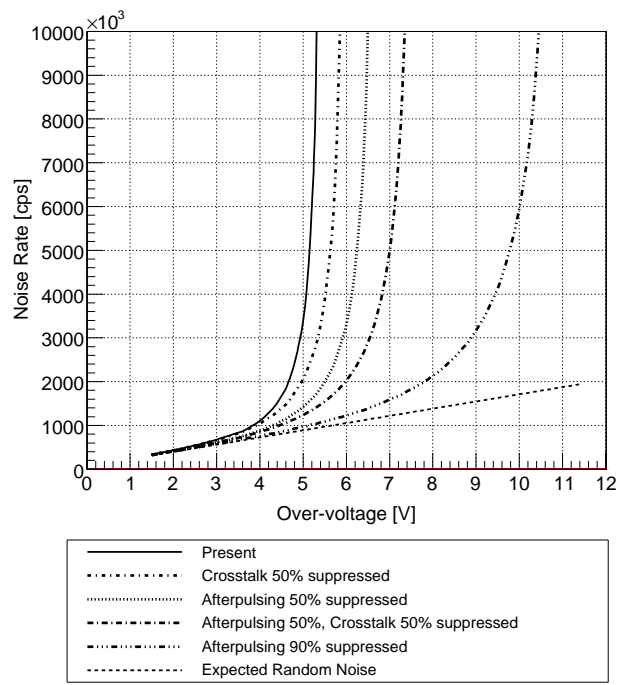}
\caption{The expected noise rate with various after-pulsing and crosstalk probabilities.}
\label{reconst_ext}
\end{center}
\end{figure}

We carried out calculations of the total-noise rate for the following cases: (i) After-pulsing probability 50\% reduced. (ii) Crosstalk probability 50\% reduced. (iii) Both after-pulsing and crosstalk probability 50\% reduced. (iv) After-pulsing 90\% reduced.
The results are shown in Fig.\ref{reconst_ext}.
Note that the derived formula was checked only for $1.5{\rm V} < \Delta V < 5.1{\rm V}$ thus that the results of the calculation for $\Delta V>5.1{\rm V}$ are an extrapolation.

The result shows that a 50\% reduction in crosstalk only pushes up the divergent point of the noise rate by $\sim0.5{\rm V}$.
A similar change is seen for the case of 50\% reduction in after-pulsing.
From Eq.(\ref{totnoise}) it is obvious that the $\Delta V$-dependence of the after-pulsing probability is the cause of the divergence of the noise rate.
If the after-pulsing probability is by 90\% reduced, then the increase of the noise rate becomes more gentle and the gain could be increased by a factor of $\sim2$.
According to the discussion in the previous subsections, reduction of both after-pulsing and random pulsing could be done at once by slimming the drift layer and/or the substrate.
However, a 90\% reduction of after-pulsing would be difficult since it is generated by defects in the Si lattice.

This test calculation shows that another 10\% reduction of after-pulsing and/or crosstalk will not result in a considerable increase of gain.
Even so, efforts to reduce such noises should be continued, as influences of after-pulsing and crosstalk is desired as less as possible for counting the number of photon precisely.

\subsection{Summary of discussions}
Summarizing all of the above discussions, what we found out about the 1600px MPPC are:
\begin{itemize}
\item Random-noise rate increases linearly with $\Delta V$ (see Fig.\ref{noise rate}).
\item PDE tends to saturate as $\Delta V$ increases (see Fig.\ref{pde}).
\item After-pulsing probability increases proportionally to $\Delta V^{2}$ (see Fig.\ref{afterpulse}).
\item Crosstalk probability increases drastically with $\Delta V$, but the dependency is more complex (see Fig.\ref{crosstalk}).
\item The drastic increase of total-noise rate with $\Delta V$ is due to the multiplication effect between after-pulsing and crosstalk (see Fig.\ref{reconst}).
\item After-pulsing is the main contributer to the divergence of the noise rate with $\Delta V$.
\end{itemize}

Speculative explanations of these results are summarized below:
\begin{itemize}
\item The increase of the PDE corresponds to the avalanche probability originating from electrons, while the increase of the random-noise rate corresponds to the avalanche probability originating from holes.
\item The main source of random noises is the drift layer below the multiplication region, and/or the substrate.
\item p-on-n type PPDs have a favorable structure for high photon detection efficiency while noise suppression.
\item The source of after-pulsing is also the drift layer.
\end{itemize}

\section{Proposal of a new structure}
Thus far, the causes of each noise and their multiplication effects on each other have been clarified.
These results of this study could be used in the development of a PPD structure which suppresses noise further.

Efforts to reduce the after-pulsing and crosstalk probabilities should be continued, but in order to increase the gain of PPDs, some alternative path for development is needed.
The design of a PPD is limited only by the requirement that it quenches.

\begin{figure}[tbp]
\begin{center}
\includegraphics[width=80mm]{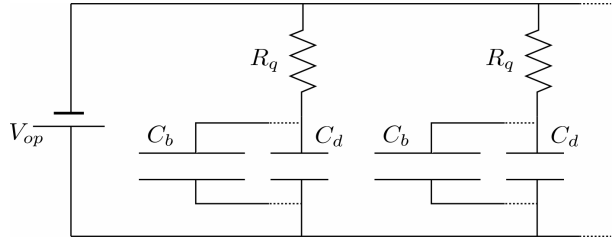}
\caption{The equivalent circuit of the new structure PPD.}
\label{new_circuit}
\end{center}
\end{figure}

Here we summarize a new proposal for a new structure of PPD which would realize high gain compatible while suppressing noises:
\begin{enumerate}
\item Add a buffer capacitor $C_{\rm b}$ parallel to the diode of the pixel (see Fig.\ref{new_circuit}).
\item Layer $C_{\rm b}$ below the photosensitive diode.
\item Use a quenching resistor with a lower resistance.
\item Shrink the drift layer and/or the substrate.
\end{enumerate}

\begin{figure}
\begin{center}
\includegraphics[width=80mm]{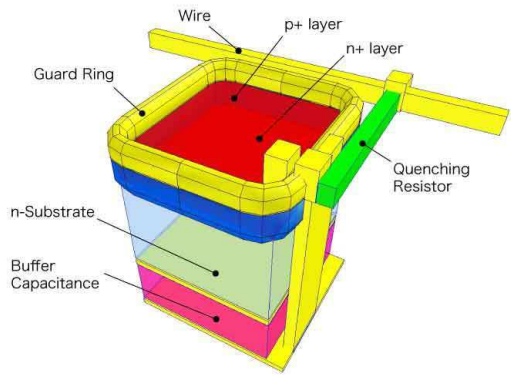}
\caption{The proposed structure of PPD.}
\label{new_structure}
\end{center}
\end{figure}

In the following the justification for each proposal is given:

The first point is the most important. By including a capacitance parallel to the diode, charges generated by avalanche multiplication in the diode are compensated for by charges flowing from the buffer capacitor, thus reducing the rate of descent of the voltage in the diode.
As a result, the multiplication process elongates and a high gain is realized before quench occurs.
Quantitatively it is expected that the gain $G$ reaches $G=(C_{\rm d}+C_{\rm b})\Delta V/e$, where $C_{\rm d}$ denotes the capacitance of the diode and $C_{\rm b}$ that of the buffer capacitor.
This expectation is also confirmed by numerical calculation.

The goal of this method is to increase the capacitance of each pixel, which in turn increases the gain.
This can be accomplished by increasing the pixel area.
However, increasing the area of the pixel also increases the volume of the depletion layer, which results in an increase in the noise rate.
Our proposal is to arrange a capacitor made of insulator parallel to the diode.
This will result in an increase in the net capacitance without changing the dimensions of the depletion layer.
Especially in the development of a PPD with high granularity, the capacitance per each pixel is small thus so is its gain.
For such a case this structure is ideal.
As a buffer capacitor high permittivity materials such as ${\rm HfO_{2}}$ are suitable.

The second point is important in order not to decrease the fill factor of the PPD.
It is expected that the technical difficulty of arranging the buffer capacitor under the diode is rather high,
especially in connecting the diode to the buffer, since it would be necessary to dig a deep pit to pass the wire through to connect them.

The third point is a countermeasure against an increase in PPD pulse time.
The role of the quenching resistor is to suppress the charges flowing into the diode from the power supply.
With a buffer capacitance inserted, since a part of the current from the quenching resistor flows into the buffer, the charges flowing into the diode are reduced.
Thus inserting a buffer enables us to use smaller (or shorter) quenching resistance.
The time constant $\tau$ of the pulses of a PPD is $\tau=R_{\rm q}(C_{\rm d}+C_{\rm b})$, where $R_{\rm q}$ denotes the resistance of the quenching resistor.
Thus inserting the buffer results in an increase of the time constant of the pulse.
However, according to the discussion above, it is possible to use smaller $R_{\rm q}$ with an increase in $C_{\rm b}$.
This implies that the time constant of the pulse could be maintained.
One of the merits of smaller (shorter) quenching resistance is the improvement of the fill factor.

The last point is intended to reduce the origin of the noise, especially random pulsing and after-pulsing.
As discussed above, it is considered that random pulsing and after-pulsing originate mainly below the multiplication layer such as in the drift layer or the substrate.
Both the thermal excitation rate and the number of defects which trap carriers are proportional to the volume of this region.
Thus reduction in the size of these region will reduce these noises.

Fig.\ref{new_structure} shows a mock-up of a PPD with the new structure.
The electrode below the substrate and another electrode below the electrode build the buffer capacitance, and the deepest electrode and the top surface of the diode are connected with a wire.
Production and development of actual devices are desired to verify that the performance expected here will be realized.

\section{Conclusion}
We have measured each type of PPD noise separately.
The drastic increase in the noise rate of the PPD with $\Delta V$ is found to be due to the multiplication effect of after-pulsing and crosstalk.
The main cause of the increase of noises is after-pulsing.
We simulated the avalanche probability originating from both electrons and holes.
It is considered that a photoelectric signals are mainly generated from electrons and random-noises are mainly generated from holes.
Based on the pulsing mechanisms of the various noises, we proposed a new structure for a PPD where a buffer capacitance is arranged in parallel to each photodiode. This has the potential to accomplish high gain with a low noise rate.

\section*{Acknowledgments}
The authors wish to express our deep appreciation to the Hamamatsu Photonics K.K. and the KEK-DTP photon-sensor group members for their
helpful discussions and suggestions.
This work was supported by Grant-in-Aid for Exploratory Research 20654021 from the Japan Society for the Promotion of Science (JSPS) and by Grant-in-Aid for JSPS Fellows 20.4439.


\begin{thebibliography}{99}
\bibitem{Renker} D.~Renker, ``Geiger-mode avalanche photodiodes, history, properties and problems'' Nucl. Instrum. Meth. A 567 (2006) 48-56
\bibitem{Otono} H.~Otono {\it et al.}, ``Study of the internal mechanisms of Pixelized Photon Detectors operated in Geiger-mode'', arXiv:0808.2541
\bibitem{Buzhan} P.~Buzhan {\it et al.}, ``Large are silicon photomultipliers: Performance and applications'', Nucl. Instrum. Meth. A 567 (2006) 78-82
\bibitem{Sciacca} E.~Sciacca {\it et al.}, ``Crosstalk Characterization in Geiger-Mode Avalanche Photodiode Arrays'', IEEE Elec. Dev. Lett., Vol. 29, no. 3, 218-220
\bibitem{Rech} I.~Rech {\it et al.}, ``A New Approach to Optical Crosstalk Modeling in Single-Photon Avalanche Diodes'', IEEE Photon. Tech. Lett. Vol. 20, no. 5, 330-332
\bibitem{HPK} K.~Yamamoto {\it et al.}, ``Newly Developed Semiconductor Detectors'', in proceedings of PD07, PoS(PD07)004
\bibitem{MPPC} S.~Uozumi {\it et al.}, ``Development and study of the multi pixel photon counter'', Nucl. Instrum. Meth. A 581 (2008) 427-432
\bibitem{Cova} S.~Cova, G. Ripamonti, ``Trapping Phenomena in Avalanche Photodiodes on Nanosecond Scale'', IEEE Elec. Dev. Lett., 12 (12) 1991
\bibitem{NDIP} H.~Oide, {\it et al.}, ``On the basic mechanism of PPDs'', arXiv:0808.2546
\bibitem{Retiere} F.~Reti\`ere, {\it et al.}, ``Using MPPCs for T2K Fine Grain Detector'', in proceedings of PD07, PoS(PD07)017
\bibitem{Uozumi} S.~Uozumi, {\it et al.}, ``Development and study of the multi pixel photon counter'', NIMA 581 (2007) 427-432
\bibitem{Oide} H.~Oide, {\it et al.}, ``Study of afterpulsing of MPPC with waveform analysis'', in proceedings of PD07, PoS(PD07)008
\bibitem{Grundmann} M.~Grundmann, ``The Physics of Semiconductors'', Springer (2006)
\bibitem{II} K.~De~Meyer {\it et al.}, ``Impact ionization in silicon: a review and update'', Solid-State Elec. Vol.33, no. 6, (1990) 705-718
\bibitem{Dash} W.C.~Dash, {\it et al.}, ``Intrinsic Optical Absorption in Single-Crystal Germanium and Silicon at ${\rm 77^{\circ}K}$ and ${\rm 300^{\circ}K}$'', Phys. Rev. 99, 1151, 1955
\bibitem{Piemonte} C.~Piemonte, ``A new Silicon Photomultiplier structure for blue light detection'', Nucl. Instrum. Meth. A 568 (2006) 224-232
\end{thebibliography}
\end{document}